\documentclass[preprintnumbers,aps,pra,groupedaddress]{revtex4}

\begin{document}
\draft
\title{Quantum coherence between mass eigenstates of a neutrino can be
destroyed by its mass-momentum entanglement}
\author{Shi-Biao Zheng}
\thanks{E-mail: t96034@fzu.edu.cn}
\address{College of Physics and Information Engineering, Fuzhou University,\\
Fuzhou 350108, China}

\begin{abstract}
If a neutrino or antineutrino produced in the decay of an unstable particle
is not entangled to its accompanying particles, its mass is necessarily
correlated with its momentum. In this manuscript, I illustrate that this
entanglement would destroy the quantum coherence between the neutrino's mass
eigenstates in both the momentum and position representations, which was
overlooked by other authors in previous investigations of entanglement and
coherence associated with neutrino oscillations. I further point out that
the states of a neutrino and an electron become nonseparable after their
charged-current interaction. This nonseparability leads to decoherence for
neutrinos propagating in matter, but was not taken into consideration in
previous investigations of the matter effect.
\end{abstract}

\vskip0.5cm

\narrowtext

\maketitle
\bigskip To interpret flavor transformations of neutrinos, each of the three
flavor eigenstates is assumed to be formed by a coherent superposition of
three mass eigenstates [1-3]. Due to the time-evolving phase differences
among the probability amplitudes associated with these eigenstates, an
initial flavor eigenstate will evolve into a superposition of all the three
flavor eigenstates, whose populations oscillate with time. In a recent
manuscript [4], I proved that the electron antineutrino produced from the $%
\beta $ decay of a neutron cannot be in a coherent superposition of
different mass eigenstates, as a consequence of their correlations with
different joint momentum states of the antineutrino and the accompanying
particles, i.e., the electron and proton. This result somehow contradicts
with the conclusions drawn in other papers [5-7], where the decoherence
caused by the entanglement between the neutrino and the accompanying
particles was also correctly recognized, but oscillations were still
predicted. As detailed below, this inconsistency is due to the fact that the
entanglement between the mass and momentum of the neutrino can also destroy
the coherence among its mass eigenstates, which was not taken into
consideration in these investigations.

The state of the entire proton-electron-antineutrino system, produced by the 
$\beta $ decay, can be written as%
\begin{equation}
\left\vert \psi \right\rangle =%
\mathop{\displaystyle\sum}%
\limits_{j}\int_{{\bf \sigma }_{j}}d^{3}{\bf P}_{\nu ,j}d^{3}{\bf P}%
_{p,j}d^{3}{\bf P}_{e,j}F({\bf P}_{\nu ,j},{\bf P}_{p,j},{\bf P}%
_{e,j})\left\vert \stackrel{-}{\nu }_{j}\right\rangle \left\vert {\bf P}%
_{\nu ,j},{\bf P}_{n,j},{\bf P}_{e,j}\right\rangle ,
\end{equation}%
where $\left\vert {\bf P}_{\nu ,j}\right\rangle $, $\left\vert {\bf P}%
_{p,j}\right\rangle $, and $\left\vert {\bf P}_{e,j}\right\rangle $
respectively denote the momentum eigenstates of the antineutrino, proton,
and electron, and ${\bf \sigma }_{j}$ denotes the distribution region of the
joint antineutrino-proton-electron momentum associated with the antineutrino
mass eigenstate $\left\vert \stackrel{-}{\nu }_{j}\right\rangle $. As proved
in Ref. [4], there is no overlapping between the momentum distribution
regions associated with different antineutrino mass eigenstates, that is, $%
{\bf \sigma }_{j}\cap $ ${\bf \sigma }_{k}=\oslash $ for $j\neq k$. With a
suitable momentum distribution of the neutron undergoing the $\beta $ decay,
the antineutrino can be disentangled with the accompanying particles. In
this case, the antineutrino's mass is necessarily correlated with its
momentum, described by the entangled state 
\begin{equation}
\left\vert \varphi _{\nu }\right\rangle =%
\mathop{\displaystyle\sum}%
\limits_{j}\int_{{\bf \sigma }_{j}}d^{3}{\bf P}_{\nu ,j}G({\bf P}_{\nu
,j})\left\vert \stackrel{-}{\nu }_{j}\right\rangle \left\vert {\bf P}_{\nu
,j}\right\rangle ,
\end{equation}%
where ${\bf P}_{\nu ,j}\neq {\bf P}_{\nu ,k}$ for $j\neq k$. When the
antineutrino's momentum degree of freedom is traced out, the mass degree of
freedom is in a classical mixture, described by the density operator 
\begin{equation}
\rho _{\nu }=%
\mathop{\displaystyle\sum}%
\limits_{j}D_{j}\left\vert \stackrel{-}{\nu }_{j}\right\rangle \left\langle 
\stackrel{-}{\nu }_{j}\right\vert ,
\end{equation}%
where $D_{j}=\int_{{\bf \sigma }_{j}}d^{3}{\bf P}_{\nu ,j}\left\vert G({\bf P%
}_{\nu ,j})\right\vert ^{2}$. This implies that the entanglement with the
momentum destroys the coherence among the mass eigenstates, prohibiting
occurrence of flavor oscillations, which can be interpreted in terms of
complementarity [8-15]. The information about the mass eigenstate of the
antineutrino is encoded in its momentum. The fact that which eigenstate the
antineutrino is in can be determined by measuring its momentum in principle
is sufficient to destroy the coherence among the mass eigenstates. This
decoherence has been overlooked in previous investigations of entanglement
and coherence in neutrino oscillations by other authors [5-7].

This decoherence can also be illustrated in the position representation, in
which the evolution of the wave function is given by%
\begin{equation}
\left\vert \varphi _{\nu }(t)\right\rangle =%
\mathop{\displaystyle\sum}%
\limits_{j}\int d^{3}{\bf r}f_{j}({\bf r},t)\left\vert \stackrel{-}{\nu }%
_{j}\right\rangle \left\vert {\bf r}\right\rangle ,
\end{equation}%
where 
\begin{equation}
f_{j}({\bf r},t)=(2\pi )^{-3/2}\int_{{\bf \sigma }_{j}}d^{3}{\bf P}_{\nu
,j}G({\bf P}_{\nu ,j})e^{i({\bf P}_{\nu ,j}\cdot {\bf r-}E_{j}t)},
\end{equation}%
$E_{j}=\sqrt{p_{j}^{2}+m_{j}^{2}}$ with $m_{j}$ being the mass of the $j$th
mass eigenstate, and $\left\vert {\bf r}\right\rangle $ denotes the position
eigenstate. The coherence between $\left\vert \stackrel{-}{\nu }%
_{j}\right\rangle $ and $\left\vert \stackrel{-}{\nu }_{k}\right\rangle $
manifested on the detection of the antineutrino is given by 
\begin{eqnarray}
C_{j,k} &=&\int_{D}d^{3}{\bf r}f_{j}^{\ast }({\bf r},t)f_{k}({\bf r},t) 
\nonumber \\
&=&(2\pi )^{-3}\int_{{\bf \sigma }_{j}}d^{3}{\bf P}_{\nu ,j}\int_{{\bf %
\sigma }_{k}}d^{3}{\bf P}_{\nu ,k}G^{\ast }({\bf P}_{\nu ,j})G({\bf P}_{\nu
,k})\int_{D}d^{3}{\bf r}e^{i[({\bf P}_{\nu ,k}-{\bf P}_{\nu ,j})\cdot {\bf %
r-(}E_{k}-E_{j})t]},
\end{eqnarray}%
where $D$ is the detection region of the antineutrino. When the size of the
detector is much larger than that of the antineutrino's wavepacket, $%
\int_{D}d^{3}{\bf r}e^{i({\bf P}_{\nu ,j}-{\bf P}_{\nu ,k})\cdot {\bf r}}$
can be well approximated by taking the integral over the whole space. As $%
{\bf P}_{\nu ,j}\neq {\bf P}_{\nu ,k}$, such an integral is zero, which
implies that the coherence $C_{j,k}$ vanishes. This result can also be
understood in terms of the position-dependent phase difference between $%
\left\vert \stackrel{-}{\nu }_{j}\right\rangle $ and $\left\vert \stackrel{-}%
{\nu }_{k}\right\rangle $ owing to the associated momentum difference. We
note that such phase differences were also included in previous
investigations, exemplified by the statement {\sl "Since }$p_{x,i}\neq
p_{x,j}${\sl , phase differences exist between the components at the point
of detection}" in Ref. [6]. However, the statement "{\sl As a result, the
interference effects of neutrino oscillations arise solely from the
different momenta in the components in the final state}" is incorrect.
Actually, the interference effects of mass eigenstates (internal degree of
freedom) would be averaged out when integrating the position (external
degree of freedom) over the large volume of the detector, as the
position-averaged value of the phase factor caused by the corresponding
momentum difference is zero.

Therefore, if the state of the electron antineutrino produced from the $%
\beta $ decay of a neutron consists of three mass eigenstates, the mass
degree of freedom of the antineutrino is necessarily entangled with the
momentum degrees of freedom of the accompanying particles or/and that of the
antineutrino itself. This conclusion holds for the neutrino or antineutrino
produced in the weak charged-current decay of other unstable particles,
including mesons and muons. It is also applicable to solar $^{8}$B
neutrinos, which are produced by the reaction [3] 
\begin{equation}
^{8}\text{B}\rightarrow ^{8}\text{Be}+\text{e}^{+}+\nu _{e}.
\end{equation}%
I further note that even if solar $^{8}$B neutrinos can be initially in a
superposition of mass eigenstates, they cannot adiabatically evolve into a
pure mass eigenstate, where the population of the electron flavor eigenstate
was assumed to be about 1/3 [3], as will be detailedly interpreted below.

Previously, the flavor transformation of solar $^{8}$B neutrinos was
attributed to the matter effect proposed by Mikheyev and Smirnov by
extending the idea of Wolfenstein, referred to as the MSW effect [16-18].
This effect originates from the charged-current (CC) interaction between the
electron neutrino and the background electrons in matter, which can be
described by the effective Hamiltonian, 
\begin{equation}
H_{cc}=\frac{G_{F}}{\sqrt{2}}{\bf \nu }_{e}^{+}\gamma _{4}\gamma _{\lambda
}(1+\gamma _{5}){\bf ee}^{\dagger }\gamma _{4}\gamma _{\lambda }(1+\gamma
_{5}){\bf \nu }_{e},
\end{equation}%
where ${\bf \nu }_{e}$ and ${\bf e}$ denote the fields associated with the
electron neutrino and electron, respectively. Using the Fierz
transformation, the Hamiltonian was rewritten in the form of%
\begin{equation}
H_{cc}^{\prime }=\frac{G_{F}}{\sqrt{2}}{\bf \nu }_{e}^{+}\gamma _{4}\gamma
_{\lambda }(1+\gamma _{5}){\bf \nu }_{e}{\bf e}^{\dagger }\gamma _{4}\gamma
_{\lambda }(1+\gamma _{5}){\bf e}.
\end{equation}%
Then the electron field was considered as a static background, whose state
is not affected by the CC interaction, so that ${\bf e}^{\dagger }\gamma
_{4}\gamma _{\lambda }(1+\gamma _{5}){\bf e}$ can be replaced with $\delta
_{\lambda ,4}N_{e}$, where $N_{e}$ is the number density of electrons. With
this treatment, the Hamiltonian $H_{cc}^{\prime }$ is effectively equivalent
to an external potential for the neutrino, given by $V=\sqrt{2}N_{e}G_{F}$.

The matter effect was interpreted in terms of the postulation that each
neutrino flavor eigenstate is formed by a linear superposition of three mass
eigenstates 
\begin{equation}
\left\vert \nu _{\alpha }\right\rangle =%
\mathop{\displaystyle\sum}%
\limits_{j=1}^{3}U_{\alpha j}\left\vert \nu _{j}\right\rangle ,
\end{equation}%
where $j$ labels the mass eigenstate, and $\alpha =e,\mu ,\tau $ denotes the
flavor of the neutrino. $\left\vert \nu _{e}\right\rangle $ was supposed to
be approximated by a superposition of $\left\vert \nu _{1}\right\rangle $
and $\left\vert \nu _{2}\right\rangle $, i.e., $U_{e3}\simeq 0$ [3]. Under
this assumption, neither $H_{cc}^{\prime }$ nor the free Hamiltonian can
couple $\left\vert \nu _{e}\right\rangle $ to $\left\vert \nu
_{3}\right\rangle $, and thus the population of $\left\vert \nu
_{3}\right\rangle $ can be neglected for the initial state $\left\vert \nu
_{e}\right\rangle $. Then the dynamics can be described in a two-dimensional
subspace $\{\left\vert \nu _{e}\right\rangle ,\left\vert \nu _{\beta
}\right\rangle \}$, where%
\begin{equation}
\left\vert \nu _{\beta }\right\rangle ={\cal N}_{\beta }(U_{\tau
3}\left\vert \nu _{\mu }\right\rangle -U_{\mu 3}\left\vert \nu _{\tau
}\right\rangle ),
\end{equation}%
with ${\cal N}_{\beta }=\left( \left\vert U_{\tau 3}\right\vert
^{2}+\left\vert U_{\mu 3}\right\vert ^{2}\right) ^{-1/2}$. Within this
subspace, the Hamiltonian can be approximately expressed as%
\begin{equation}
H\simeq V\left\vert \nu _{e}\right\rangle \left\langle \nu _{e}\right\vert +%
\mathop{\displaystyle\sum}%
\limits_{\eta ,\xi =e,\beta }M_{\eta ,\xi }(p)\left\vert \nu _{\eta
}\right\rangle \left\langle \nu _{\xi }\right\vert ,
\end{equation}%
where%
\begin{equation}
M_{\eta ,\xi }(p)\simeq 
\mathop{\displaystyle\sum}%
\limits_{j=1}\frac{m_{j}^{2}}{2p}U_{\eta j}^{\ast }U_{\xi j},
\end{equation}%
with $U_{\beta j}={\cal N}_{\beta }(U_{\tau 3}U_{\mu j}-U_{\mu 3}U_{\tau j})$%
. Here $p$ denotes the neutrino momentum, which is much larger than the mass
($m_{j}$) associated with each mass eigenstate. The trivial common energy,
described by $p%
\mathop{\displaystyle\sum}%
\limits_{j}\left\vert \nu _{j}\right\rangle \left\langle \nu _{j}\right\vert 
$, has been discarded. When $V\gg m_{j}^{2}/2p$, the electron flavor
approximately coincides with the eigenstate of the Hamiltonian with the
larger eigenenergy. If the electron number density is changed sufficiently
slowly, the neutrino adiabatically follows the corresponding Hamiltonian
eigenstate during its propagation. On the solar surface, $V$ can be
neglected as compared to $m_{j}^{2}/2p$ so that the eigenstates of the
Hamiltonian coincide with the mass eigenstates. This implies that the
initial electron flavor eigenstate evolves to the mass eigenstate with the
larger mass ($\left\vert \nu _{2}\right\rangle $) when the neutrino reaches
the solar surface. This mass eigenstate remains invariant until being
detected on the Earth. The resulting probability $P_{\left\vert \nu
_{e}\right\rangle \rightarrow \left\vert \nu _{e}\right\rangle }$ is
approximately equal to $\left\vert U_{2e}^{\dagger }\right\vert ^{2}$, which
was assumed to be about $1/3$ [3].

This treatment has overlooked the crucial fact that the CC reaction leads to
neutrino-electron entanglement when the neutrino is in a superposition of
the electron flavor eigenstate and the other two flavor eigenstates before
the reaction [19]. As the neutrino and the electron can be transformed into
each other by their CC reaction, it helps to make the presentation more
clear to refer the original neutrino and the original electron to as
particle 1 and particle 2, respectively. The CC reaction transforms the
state $\left\vert \nu _{e},{\bf p}_{1}\right\rangle _{1}\left\vert e,{\bf p}%
_{2}\right\rangle _{2}$ into $\left\vert e,{\bf p}_{3}\right\rangle
_{1}\left\vert \nu _{e},{\bf p}_{4}\right\rangle _{2}$, where the subscripts
"1" and "2" outside the kets label the two particles, and ${\bf p}_{1}$ ($%
{\bf p}_{3}$) and ${\bf p}_{2}$ (${\bf p}_{4}$) are their momenta before
(after) the reaction. If particle 1 is initially in the flavor eigenstate $%
\left\vert \nu _{e}\right\rangle $ and ${\bf p}_{1}={\bf p}_{4}$, the Fierz
rearranging is equivalent to relabelling the two particles, which does not
cause any problem. However, when it is initially in a superposition of $%
\left\vert \nu _{e}\right\rangle $ and $\left\vert \nu _{\beta
}\right\rangle $, it will be entangled with particle 2 by the CC reaction.
To illustrate this point, we suppose that the two-particle system is
initially in the state 
\begin{equation}
\left\vert \psi _{0}\right\rangle =(C_{e}\left\vert \nu _{e},{\bf p}%
_{1}\right\rangle _{1}+C_{\beta }\left\vert \nu _{\beta },{\bf p}%
_{1}\right\rangle _{1})\left\vert e,{\bf p}_{2}\right\rangle _{2}.
\end{equation}%
In this case, the CC reaction actually corresponds to a conditional
dynamics, by which particle 1 exchanges its state with particle 2 when it is
initially in the electron flavor eigenstate, but nothing occurs if it is
initially in the other two flavor eigenstates. This conditional state
swapping evolves the system to the entangled state 
\begin{equation}
\left\vert \psi \right\rangle =C_{e}\left\vert e,{\bf p}_{3}\right\rangle
_{1}\left\vert \nu _{e},{\bf p}_{4}\right\rangle _{2}+C_{\beta }\left\vert
\nu _{\beta },{\bf p}_{1}\right\rangle _{1}\left\vert e,{\bf p}%
_{2}\right\rangle _{2}.
\end{equation}%
It should be noted that the electron transformed from the neutrino does not
have the same momentum as the original electron, i.e., ${\bf p}_{3}\neq {\bf %
p}_{2}$. Such momentum differences have been used to identify
neutrino-electron scattering events in SNO experiments [3]. This quantum
entanglement is masked by the Fierz rearranging and the subsequent
replacement of the electron part in the Hamiltonian with a number. We
further note that the Fierz rearranging is valid for calculation of the $e$-$%
\nu _{e}$ scattering amplitude, which is irrelevant to the quantum coherence
between $\left\vert \nu _{e}\right\rangle $ and $\left\vert \nu _{\beta
}\right\rangle $. However, it overlooks the fact that $\left\vert \nu
_{e}\right\rangle $ and $\left\vert \nu _{\beta }\right\rangle $ are carried
by different particles after the CC reaction, which is essential for correct
description of the neutrino state evolution in matter. In other words, the
states of the two particles are no longer separable after their CC
interaction, so that the electrons participating in such interactions cannot
be treated as a static background for the neutrino, and their effects cannot
be modeled as a potential, which cannot reflect the effect of quantum
entanglement produced by the conditional dynamics.

Due to the quantum entanglement, each of the two particles is essentially in
a mixture of the neutrino and electron states. This critical point can be
illustrated more clearly by the reduced density operators for these
particles, each obtained by tracing out the degree of freedom of the other
particle, given by%
\begin{eqnarray}
\rho _{1} &=&Tr_{2}(\left\vert \psi \right\rangle \left\langle \psi
\right\vert )  \nonumber \\
&=&\left\vert C_{e}\right\vert ^{2}\left\vert e,{\bf p}_{3}\right\rangle
_{1}\left\langle e,{\bf p}_{3}\right\vert +\left\vert C_{\beta }\right\vert
^{2}\left\vert \nu _{\beta },{\bf p}_{1}\right\rangle _{1}\left\langle \nu
_{\beta },{\bf p}_{1}\right\vert ,  \nonumber \\
\rho _{2} &=&Tr_{1}(\left\vert \psi \right\rangle \left\langle \psi
\right\vert )  \nonumber \\
&=&\left\vert C_{e}\right\vert ^{2}\left\vert \nu _{e},{\bf p}%
_{4}\right\rangle _{2}\left\langle \nu _{e},{\bf p}_{4}\right\vert
+\left\vert C_{\beta }\right\vert ^{2}\left\vert e,{\bf p}_{2}\right\rangle
_{2}\left\langle e,{\bf p}_{2}\right\vert .
\end{eqnarray}%
Under the subsequent free Hamiltonian dynamics, $\rho _{1}$ and $\rho _{2}$
evolve as%
\begin{eqnarray}
\rho _{1}^{\prime } &=&\left\vert C_{e}\right\vert ^{2}\left\vert e,{\bf p}%
_{3}\right\rangle _{1}\left\langle e,{\bf p}_{3}\right\vert +\left\vert
C_{\beta }\right\vert ^{2}\left\vert \varphi _{1},,{\bf p}_{1}\right\rangle
_{1}\left\langle \varphi _{1},{\bf p}_{1}\right\vert ,  \nonumber \\
\rho _{2}^{\prime } &=&\left\vert C_{e}\right\vert ^{2}\left\vert \varphi
_{2},{\bf p}_{4}\right\rangle _{2}\left\langle \varphi _{2},{\bf p}%
_{4}\right\vert +\left\vert C_{\beta }\right\vert ^{2}\left\vert e,{\bf p}%
_{2}\right\rangle _{2}\left\langle e,{\bf p}_{2}\right\vert ,
\end{eqnarray}%
where%
\begin{eqnarray}
\left\vert \varphi _{1}\right\rangle _{1} &=&u_{p_{1}}\left\vert \nu _{\beta
}\right\rangle _{1}+v_{p_{1}}\left\vert \nu _{e}\right\rangle _{1}, 
\nonumber \\
\left\vert \varphi _{2}\right\rangle _{2} &=&u_{p_{4}}^{\ast }\left\vert \nu
_{e}\right\rangle -v_{p_{4}}^{\ast }\left\vert \nu _{\beta }\right\rangle .
\end{eqnarray}%
$u$ and $v$ depend on time as 
\begin{eqnarray}
u_{p} &=&\cos (\lambda _{p}t)-i\frac{\Delta _{p}}{\sqrt{\lambda
_{p}^{2}+\Delta _{p}^{2}}}\sin (\lambda _{p}t),  \nonumber \\
v_{p} &=&\frac{-i\lambda _{p}}{\sqrt{\lambda _{p}^{2}+\Delta _{p}^{2}}}%
e^{i\theta _{p}}\sin (\lambda _{p}t),
\end{eqnarray}%
where $\Delta _{p}=[M_{e,e}(p)-M_{\beta ,\beta }(p)]/2$, $\lambda
_{p}=\left\vert M_{e,\beta }(p)\right\vert $, and $\theta _{p}=\arg
[M_{e,\beta }(p)]$. After this free evolution, the total $\left\vert \nu
_{e}\right\rangle $-state population is

\begin{eqnarray}
P_{\left\vert \nu _{e}\right\rangle } &=&\int d^{3}{\bf p(}_{1}\left\langle
\nu _{e},{\bf p}\right\vert \rho _{1}^{\prime }\left\vert \nu _{e},{\bf p}%
\right\rangle _{1}+_{2}\left\langle \nu _{e},{\bf p}\right\vert \rho
_{2}^{\prime }\left\vert \nu _{e},{\bf p}\right\rangle _{2})  \nonumber \\
=\left\vert C_{e}u_{p_{4}}\right\vert ^{2} &&+\left\vert C_{\beta
}v_{p_{1}}\right\vert ^{2}.
\end{eqnarray}%
Such a probability does not present the cross terms proportional to $%
C_{e}^{\ast }C_{\beta }$ and $C_{e}C_{\beta }^{\ast }$. This is due to the
fact that the state components $\left\vert \varphi _{1}\right\rangle _{1}$
and $\left\vert \varphi _{2}\right\rangle _{2}$ have different momenta and
are carried by different particles, so that quantum interference cannot
occur.

If one only concerns about the neutrino part in the two-particle system, its
behavior just after the CC reaction can be effectively described by the
classically mixed state%
\begin{equation}
\rho _{\nu }=\left\vert C_{e}\right\vert ^{2}\left\vert \nu _{e},{\bf p}%
_{4}\right\rangle \left\langle \nu _{e},{\bf p}_{4}\right\vert +\left\vert
C_{\beta }\right\vert ^{2}\left\vert \nu _{\beta },{\bf p}_{1}\right\rangle
\left\langle \nu _{\beta ,},{\bf p}_{1}\right\vert .
\end{equation}%
However, it should be born in mind that the two mixed state components are
essentially carried by two different particles. The validity of this
description can be illustrated by examining the subsequent free Hamiltonian
dynamics, which evolves $\rho _{\nu }$ to%
\begin{equation}
\rho _{\nu }^{\prime }=\left\vert C_{e}\right\vert ^{2}\left\vert \varphi
_{2},{\bf p}_{4}\right\rangle \left\langle \varphi _{2},{\bf p}%
_{4}\right\vert +\left\vert C_{\beta }\right\vert ^{2}\left\vert \varphi
_{1},{\bf p}_{1}\right\rangle \left\langle \varphi _{1},{\bf p}%
_{1}\right\vert .
\end{equation}%
The resulting neutrino's $\left\vert \nu _{e}\right\rangle $-state
probability is the same as Eq. (13). This equivalence further confirms that
the CC reaction indeed destroys the quantum coherence between $\left\vert
\nu _{e}\right\rangle $ and $\left\vert \nu _{\beta }\right\rangle $. When a
second CC reaction occurs, these two particles will be further entangled
with a third particle. Under the competition between the coherent coupling
and CC-reaction-induced decoherence, the population of $\left\vert \nu
_{e}\right\rangle $ is progressively decreased while that of $\left\vert \nu
_{\beta }\right\rangle $ is increased until reaching the steady state%
\begin{equation}
(\left\vert \nu _{e}\right\rangle \left\langle \nu _{e}\right\vert
+\left\vert \nu _{\beta }\right\rangle \left\langle \nu _{\beta }\right\vert
)/2.
\end{equation}%
For simplicity, we here have discarded the momentum degrees of freedom. For
this mixed state, the gain of the $\left\vert \nu _{e}\right\rangle $-state
population originating from the $\left\vert \nu _{\beta }\right\rangle
\rightarrow \left\vert \nu _{e}\right\rangle $ transition cancels out the
loss due to the $\left\vert \nu _{e}\right\rangle \longleftrightarrow
\left\vert \nu _{\beta }\right\rangle $ transition. Therefore, the
probability $P_{\left\vert \nu _{e}\right\rangle \rightarrow \left\vert \nu
_{e}\right\rangle }$ should not be smaller than 1/2, which is inconsistent
with the solar $^{8}$B neutrino experiments [3,20].

The observed deficit of solar $^{8}$B electron neutrinos can be well
explained in terms of a new mechanism, where the neutrino can oscillate
among different flavors by interacting with the Z bosonic field [19]. If the
Z bosonic field is allowed to connect different neutrino flavors, its
virtual excitation induces the coherent couplings among these flavors, given
by the Hamiltonian 
\begin{equation}
{\cal H}=%
\mathop{\displaystyle\sum}%
\limits_{\alpha ,\beta }\xi _{\alpha \beta }\left\vert \nu _{\alpha
}\right\rangle \left\langle \nu _{\beta }\right\vert ,
\end{equation}%
where $\alpha ,\beta =e,\mu ,\tau $. On the other hand, the decoherence
induced by the CC reaction can be modeled as the incoherent transformation 
\begin{equation}
\rho \rightarrow S_{e}\rho S_{e}+S_{\mu +\nu }\rho S_{\mu +\nu },
\end{equation}%
where $S_{e}=\left\vert \nu _{e}\right\rangle \left\langle \nu
_{e}\right\vert $, $S_{\mu +\nu }=\left\vert \nu _{\mu }\right\rangle
\left\langle \nu _{\mu }\right\vert +\left\vert \nu _{\tau }\right\rangle
\left\langle \nu _{\tau }\right\vert $, and $\rho $ denotes the density
operator of the neutrino before the CC reaction. Under the competition
between the coherent coupling and decoherence effect, the neutrino has a
unique steady state, given by%
\begin{equation}
\rho _{s}=\frac{1}{3}%
\mathop{\displaystyle\sum}%
\limits_{\alpha }\left\vert \nu _{\alpha }\right\rangle \left\langle \nu
_{\alpha }\right\vert .
\end{equation}%
This steady state corresponds to the equal classical mixture of the three
flavors, which is not affected both by the coherent coupling and by the
CC-reaction-induced decoherence. This steady state is in well agreement with
the solar $^{8}$B neutrino experiments [3,20].

In summary, there necessarily exist mass-momentum entanglement for the
neutrino produced by the decay of an unstable particle, if the neutrino's
flavor eigenstates are inconsistent with mass eigenstates and the neutrino
is not entangled with the accompanying particles. Due to this entanglement,
the neutrino is essentially in a classical mixture of these mass eigenstates
when all the momentum freedom degrees are traced out. Consequently, the
neutrino cannot has a definite flavor at the production, and the population
of each flavor eigenstate cannot be changed during the propagation. For
solar $^{8}$B neutrinos, even if they can be initially in a superposition of
different mass eigenstates, they cannot adiabatically evolve into a pure
mass eigenstate, whose electron flavor population was assumed to be about
1/3. These results imply that none of the observed flavor transformations of
neutrinos or antineutrinos can be consistently interpreted in terms of the
inconsistency between the flavor and mass eigenstates. The experimental
results can be well explained within the framework where the coherent flavor
coupling is mediated by virtual excitation of the Z bosonic field, and the
CC interaction is modeled as a dephasing effect [19].

\end{document}